\documentstyle[prl,aps]{revtex}
\begin{document}
\draft

\title{Spin--Down Power in Astrophysics\footnote{This essay received an ``honorable mention'' in the 2001 Essay Competition of the Gravity Research Foundation.}}
\author{Feng Ma}

\address{Prc-Mrc 2nd Floor/R9950, \\The University of Texas at Austin, 
Austin, TX 78712, USA\\Email: feng@astro.as.utexas.edu}

\maketitle

\begin{abstract}
While the accretion
power in astrophysics has been studied in many 
astronomical environments, 
the ``spin--down power'' is often neglected. 
In this essay I demonstrate that the 
spin--down power alone may drive a rotating system from
sub-critical condition to critical condition with a small but 
finite probability. In the case of an isolated 
spinning-down neutron star, the star may undergo
a quark--hadron phase transition in its center and become observable 
as a soft gamma repeater or a cosmological gamma--ray burst.  
For a spinning--down white dwarf, its Chandrasekhar mass limit  
will decrease and may reach the stellar mass, then the star explodes to 
a type Ia supernova. 
Gravitational wave detectors may be able
to test these models.
\end{abstract}

~

Accretion power plays a fundamental role in astrophysics~\cite{frank}, 
and is believed to be the energy source for active galactic
nuclei, Galactic X-ray sources, novae and supernovae. 
What about spin--down power?
Most stars are spinning down while losing angular momentum. 
The decrease in centrifugal 
force leads to an increase in the stellar central density $\rho_c$.
$\Delta\rho_c$ is generally small unless
a star spins down from an initial frequency 
close to its Keplerian frequency $\Omega_K$
(the frequency at which the star begins to shed matter near its equator). 
The spin--down power is often neglected for two 
reasons. First, most stars spin much slower than their $\Omega_K$. 
However, as I shall show in this essay, if a star has a critical central
density or a critical mass, 
the spin--down power alone can drive the star to have 
a catastrophic transition from sub-critical condition to critical condition, 
and cause a  release of large amount of  energy in observable
forms. Second, the conventional way of
treating rotating stars misses the evolution of an 
individual star. Stars of one type are often treated
as a set, and relations such as $M_*-\rho_c$ are usually  
solved in the literature. 
Rotation leads to a ``mass increase'' in the  $M_*-\rho_c$
plot, which is actually the mass difference between 
two stars with the same $\rho_c$ at different
angular velocities. These theoretical 
results are difficult to test observationally. 
The evolution of an individual
rotating star, in terms of tracing $\rho_c$
and overall structural changes over time, has rarely been
studied. That is likely why we are missing some
interesting explanations of many astrophysical phenomena. 

To solve the structure of a rotating 
relativistic star, Hartle
developed~\cite{Hartle1,Hartle2} a perturbation
solution based on the Schwarzschild metric of a static, spherically
symmetric object. 
Rotation distorts the star away from spherical symmetry.  
By treating $\rho_c$ as an input parameter and $\Delta{M_*}$ as
a perturbation, $M_*-\rho_c$ relations 
and the ``mass increase'' due to rotation
can be solved for different rotational frequencies $\Omega$ (as seen by 
a distant observer)~\cite{Hartle2,WG,Colpi}. 
We have noted that, instead of deriving $M_*-\rho_c$ relations for a family 
of neutron stars, tracing the evolution of an individual 
star gives better insight~\cite{maluo}. 
Along with an increase in $\rho_c$, the overall structure, 
chemical composition and 
spin--down behavior of a star are modified. 
In principle, we can derive a set of equations 
parallel to those of Hartle, with $M_*$ as an input parameter and  
$\Delta\rho_c$ as the perturbation, and solve  $\rho_c$ and 
stellar structures at different $\Omega$.  
In practice, it is more convenient to exploit Hartle's method
to solve $\Delta\rho_c$ for an individual star, using the approximation 
that a star has a constant 
gravitational mass.  A clever way to do this is to  
first plot  $M_*-\rho_c$ relations
at different $\Omega$ using Hartle's 
method~\cite{Hartle2,WG,Colpi}, then cross these curves 
with a line of constant mass. The projected $\rho_c$'s
are  those  of  an individual
star at different $\Omega$s. The structural
change of a spinning--down star can thus be solved
self-consistently when the Equations of State (EOS) 
of the star are known. 

The EOS of quark matter is 
much softer than that of neutron matter because of 
the QCD asymptotic freedom. A neutron star containing
a quark core is thus more compact and has a larger $\Omega_K$
than a normal neutron star. 
If the initial spin period of a neutron star is
$P_i{\approx}20$ ms, e.g. that of the Crab pulsar, it will have a central
density increase ${\Delta}\rho_c/\rho_c{\sim}$0.1\% in its life time.
Assuming a critical density  $\rho_{\rm cr}$ for a phase transition,
only those neutron stars born with 
$\rho_{\rm cr}(1{-}\Delta\rho_c/\rho_c){<}\rho_c{<}\rho_{\rm cr}$
would have the chance to evolve from sub-$\rho_{\rm cr}$ to $\rho_{\rm cr}$. 
If $\rho_{\rm cr}$ lies between the lower and upper limits
for the central density, i.e.
$\rho_{\rm l}{<}\rho_{\rm cr}{<}\rho_{\rm u}$, 
a small but finite fraction of neutron stars will
undergo the transition at a rate ~\cite{ma96}
\begin{equation}
 \displaystyle
    R = \frac{\rho_{\rm cr}(\Delta{\rho_c}/\rho_c)}{\rho_{\rm u} - \rho_
{\rm l}}~  R_{\rm NS} \simeq 10^{-5} \left(\frac{P_i}{20{\rm\, ms}}\right)^{-2} 
\left(\frac{R_{\rm NS}}{10^{-2}}\right) {\rm yr}^{-1} {\rm galaxy}^{-1},  \label{eq:rate}
\end{equation}
where $R_{\rm NS}$ is  the neutron star birth rate. 
Clearly these events can be observed on a 
regular basis with
millions of galaxies in our view.

A catastrophic phase transition inside a spinning--down neutron star 
can happen on a  time scale of seconds. 
In such a short time, the star collapses from 
a radius of $r\sim$15 km to $r\sim$10 km, 
associated with a sudden spin--up. The gravitational energy released 
$E{\simeq}
10^{53}\Delta{r}/r$  ergs 
is large enough to power a cosmological gamma-ray burst (GRB)~\cite{ma96}. 
Observed GRBs have total 
energy of $10^{51}-10^{54}$ ergs (assuming the energy is emitted isotropically), 
which were explained with colliding neutron stars~\cite{pac86}. 
The event rate is estimated to be $10^{-5}$ yr$^{-1}$ galaxy$^{-1}$ if 
the emission is isotropic, and can be much higher if 
the emission is highly beamed. Equation (\ref{eq:rate}) 
can still account for the GRBs
even if they are beamed, as long as faster initial
spins are sought. The faster initial spin
of a neutron star naturally offers a large angular 
momentum, that helps beam the electromagnetic 
radiations.

Quark--hadron phase transition is
different from water--vapor phase transition even though
it is likely to be first order. 
It has an additional freedom of 
whether a proton deconfines to $uud$ quarks
or a neutron becomes $udd$ quarks. Hence, 
electric charges are not conserved in each of the two phases, although
overall charge neutrality is achieved through
leptons. Consequently, the two phases are not necessarily separated
by gravity~\cite{glenn92}. If this were the case, the phase transition
will happen slowly on a time scale of $10^5$ years, and 
we should be able to observe $\sim$1 event in our Galaxy 
at any moment. Gravitational energy is released at an average  rate of
$10^{40}$ ergs s$^{-1}$ during this slow phase transition. 
Most of this energy is 
released via neutrino emission, 
and only a tiny fraction is used to heat the star up to a surface 
temperature of $3{\times}10^{6}$ K, 
yielding a soft X-ray luminosity of 
${\sim}10^{35}$ ergs s$^{-1}$. This is still 25 times more luminous 
than the sun~\cite{ma98}! 
Unlike many Galactic X-ray sources powered by accretion
in binary systems, these types of X-ray sources can be isolated objects.

While the fluid core of the star is contracting, 
stress builds up in the solid crust. The
cracking of the crust releases bursts of energy
that can be observed as 
Soft Gamma Repeaters (SGRs)~\cite{duncan92}. SGRs are X-ray transient sources 
associated with young ($10^4$ yr) 
supernova remnants (SNRs). 
They are also usually quiescent X-ray emitters 
(with $kT{\sim}$ 1 keV, $L_{\rm X}{\sim}10^{35}$ ergs s$^{-1}$). 
So far four SGRs have been discovered in the Galaxy, 
and one in the Large Magellanic Cloud. 
Two of these SGRs have characteristic ages 
$\tau_c{=}\Omega/2|\dot{\Omega}|$ (${\sim}10^3$ yrs) 
$<$ SNR age (${\sim}10^4$ yrs)~\cite{kouv99}. We already
know that $\tau_c$ should be an upper limit for a pulsar's age.
How does it reconcile with the age of an SNR, which is responsible for
the birth of the pulsar in the first place? 
This can be explained easily with the picture of phase transition. 
The phase transition tends to spin--up a neutron star
while making it more compact and easier to brake, 
and hence $\tau_c$ may {\it underestimate} the true age while 
it is an {\it upper} limit for normal neutron stars. 


Now let us take a look at the role of spin--down power in
the progenitors of type Ia supernovae (SNe Ia).  
SNe Ia are believed to be explosions
of Chandrasekhar mass carbon--oxygen White Dwarfs (WDs) 
in binary systems~\cite{wheeler96}. 
However, detailed observations 
have ruled out almost any accretion rate in binary 
evolutions~\cite{woosley90,branch95}, leading SN Ia theory to a 
paradox~\cite{wheeler90}. 
Instead of accretion, the spin--down power 
 can drive the central density of a WD to the 
critical density for carbon ignition and  trigger an explosion.
The key point is that a rotating WD has a larger Chandrasekhar  mass limit 
$M_{\rm Ch}{\simeq}M_{\rm Ch, 0}(1+{3T}/{|W|})$~\cite{Shapiro},
where $M_{\rm Ch, 0}$ is for non-rotating WDs; $T$ and $W$ are 
rotational and gravitational energy, respectively. 

Observed rotation of 
WDs (most of which have masses ${\sim}0.6 M_{\odot}$) is generally 
small with $T/|W|{\lesssim}10^{-5}$~\cite{Shapiro}.  Although 
there are not enough data to give a 
mass--rotation relation for WDs, it is likely
that more massive
WDs have higher $T/|W|$ ratios for two reasons. First, they have smaller radii; 
second, they have suffered less mass loss and thus less angular momentum loss. 
We will assume that the progenitor 
of a WD loses its outer layers and leaves behind a 
rotating core (a pre-WD), which has a uniform density
before collapsing into a WD. A $1.4 M_{\odot}$ pre-WD has a radius 1.326 times
that of  a $0.6 M_{\odot}$ pre-WD, and its moment of inertia 
$I{\simeq}0.4Mr^2$ is 3 times larger. 
After the WDs are formed, more massive ones 
have smaller radii $r_{1.4M_{\odot}}{\sim}0.3 
r_{0.6M_{\odot}}$~\cite{Shapiro}. 
It is easy to see that for the collapsed WDs $T_{1.4M_{\odot}}{\sim}80T_{0.6M_{\odot}}$ 
and $W_{1.4M_{\odot}}{=}0.25W_{0.6M_{\odot}}$. Hence, 
$3T/|W|{\sim}10^{-2}$ for a $1.4 M_{\odot}$ WD. 
As a result, $M_{\rm Ch}{=}1.01M_{\rm Ch, 0} $ and is about $1.414M_{\odot}$ 
if $M_{\rm Ch, 0}{=}1.400M_{\odot}$. 

The observed mass distribution of about 200 WDs~\cite{weidemann84,finley97}
can be roughly fit with a power law mass 
function $N(M){\sim}M^{-3}$ 
with $M$ between $0.6 M_{\odot}$  and $1.3M_{\odot}$. Although no 
WDs more massive than $1.3M_{\odot}$ are observed in these samples, 
it is likely that the high mass tail of the distribution 
from $1.400M_{\odot}$ to $1.414M_{\odot}$, extrapolated from
the observed mass function, constitutes $\sim$0.1\% of the total WD 
population. 
The spin-down time scale of these WDs is ${\sim}10^9$ yr~\cite{Shapiro}, 
during which $M_{\rm Ch}$ evolves from $1.414M_{\odot}$ to $1.400M_{\odot}$. 
As a result, about 0.1\% of all WDs will have the chance to 
evolve from sub-Chandrasekhar mass to their mass limit and undergo catastrophic
events solely due to spin--down. The number of WDs in 
a galaxy is ${\sim}10^{10}$, among which $10^7$ will be 
in the high mass tail with $M_*{>}M_{\rm Ch, 0}$ 
and will undergo catastrophic events due to spin--down within
$10^9$ yr. Hence, the event rate is $10^{-2}$ yr$^{-1}$ galaxy$^{-1}$. 
The fate of these WDs is bifurcated. If their central densities reach the critical
density for carbon ignition, they will become SNe Ia. 
If not, they will tend to collapse
into black holes or neutron stars. The latter may be the origin of 
isolated millisecond pulsars. 

In binary evolution, in addition to the mass accretion, 
a WD can be spun--up. A WD can grow significantly more 
massive than $M_{\rm Ch, 0}$ 
without exploding, if the transfer 
of angular momentum is efficient. 
It is possible that a WD explodes long after 
the accretion ceases.
It is intuitive to define an ``effective accretion rate''
$\dot{M}_{\rm eff}{=}3TM_{\rm Ch, 0}/({\tau}|W|)$, 
where $\tau$ is the time scale. In the case of accretion and 
spin--up, $\dot{M}_{\rm eff}$ is negative and is associated with 
a positive real accretion rate. In the 
case of spin--down without accretion, $\dot{M}_{\rm eff}$ 
is positive and it describes how fast
$M_{\rm Ch}$ approaches the gravitational 
mass of the WD. If a
WD is spun--up to a high rotation with 
$T/|W|{\sim}0.1$, then the gravitational 
radiation rather than the viscosity dominates the 
dissipation process, and $\tau{\sim}10^3{-}10^7$ 
yr rather than $10^9$ yr~\cite{Shapiro}. Correspondingly, 
$\dot{M}_{\rm eff}{\sim}10^{-4}{-}10^{-8}M_{\odot}$
 yr$^{-1}$, which is comparable to
real mass accretion rate. 

Analysis of the non-spherical shapes of nova remnants  
suggest that WDs can have such fast rotations~\cite{fiedler80}. 
Hence, the transport of angular momentum during the accretion 
process can be very efficient, which can be easily 
understood by realizing that part of
the huge amount of {\it orbital} angular momentum 
of the binary system is converted to {\it spin} 
angular momentum of the WD during accretion. 




Traditional mass accretion models for SNe Ia predict generally 
weak gravitational waves (GWs),
because the exploding WD is slowly rotating and the 
explosion is nearly spherical. 
For the model described in this paper, it is possible 
that the progenitors of some SNe Ia are rapidly rotating WDs, 
which undergo asymmetric
explosions and produce strong GWs. The collapsing or exploding 
rapidly rotating
compact stars can produce GWs as strong as those from stellar mergers, but the 
wave forms are very different~\cite{wheeler66}. Hence, 
it is possible to test different models for GRBs and SNe Ia in future 
GW observations such as the LIGO experiment. However, it should be noted
that GWs are emitted more or less isotropically in a GRB,
while the electromagnetic radiation is likely to be highly
beamed (with a beaming factor 100--1000) in the 
angular momentum direction. LIGO may need to 
accumulate 100--1000 events before  seeing a gravitational wave signal
at the same time as a GRB. 

{\bf Acknowledgments: }
I thank J. Craig Wheeler for advice, and 
Gauri Karve and Ariane Beck for help with the manuscript.

\end{document}